\documentclass[RNAAS]{aastex631}
\usepackage{amsmath}
\usepackage{graphicx}

\begin{document}
	
	\title{Will nanodust reappear in STEREO/WAVES data?}
	%\title{Nanodust observations by wave instruments on STEREO and Cassini  when \\ the interplanetary electric field points towards solar magnetic equator}
	
	\correspondingauthor{Nicole Meyer-Vernet}
	\email{nicole.meyer@obspm.fr}
	%\email{arnaud.zaslavsky@obspm.fr}
	
	\author[0000-0001-6449-5274]{Nicole Meyer-Vernet}
	\affiliation{LIRA, Observatoire de Paris, Université PSL, Sorbonne Université, \\ Université Paris Cité, CY Cergy Paris Université, CNRS, 92190 Meudon, France}
	%\author[0000-0001-8543-9431]{Ar naud Zaslavsky}
	%\affiliation{LESIA, Observatoire de Paris, PSL Université,\\ CNRS, Sorbonne Université, Université Paris Cité, 92195 Meudon, France}
	
	\begin{abstract}
	
	Nanodust particles produced near the Sun by collisional breakup of larger grains are accelerated   in the magnetised solar wind and reach high speeds outwards of 1 AU. Vaporisation and ionisation of fast dust grains impacting a spacecraft produce voltage pulses on wave instruments that enable them to act as dust detectors. Wave instruments on STEREO and  on Cassini during its cruise phase detected a highly variable flux of fast nanodust.  Both detections took place when the orientation of the solar magnetic dipole produced an interplanetary electric field that focused nanoparticles towards the heliospheric current sheet (HCS) - a geometry that is recurring because of the  periodicity of solar activity.
	\end{abstract}

\section{Introduction}
Our solar system zodiacal cloud includes dust particles of various sizes.  Since the Lorentz force acceleration on a particle of radius $r$ is  proportional to the charge-to-mass ratio $q/m$ with $q\propto r$ \citep{man13} and $m\propto r^3$, it controls the dynamics of small particles. Nanodust gyroradii being much smaller than the scale of variation of the magnetic field $\mathbf{B}$, they are picked up by the solar wind (of velocity $\mathbf{V}$), gyrate around  field lines  and move at
\begin{equation*}
	\mathbf{v_D}=\frac{(-\mathbf{V} \times \mathbf{B})\times \mathbf{B}}{B^2}  
\end{equation*}
($E \times B$ drift) which exceeds  300 km/s farther than 1 AU,   plus a smaller vertical drift  $v_{\theta} \propto m$ \citep{nor63}. Nanoparticles created near the Sun can be trapped  within 0.2 AU  and released intermittendly when the outer boundary of the trapping region  changes due to transient magnetic field changes  \citep{cze12}.

Fast dust grains impacting a spacecraft vaporise and ionise, producing a released electric charge scaling  linearly with mass but greater than the impact speed cubed. Charge recollection  and induced voltages on the spacecraft and antennas produce  voltage pulses on wave instruments, whose detection enables them to measure dust. Such a method, pioneered when  Voyager 2 crossed Saturn's G ring   \citep{aub83, gur83}, has been  widely applied  and is currently used in the inner heliosphere onboard Parker Solar Probe \citep{pag20} and Solar Orbiter \citep{zas21}.

\section{Interplanetary nanodust observations on STEREO and Cassini}

A  nanodust particle impacting a spacecraft at 300 km/s produces a charge nearly as large  as that produced by a much slower $\beta$ dust particle. STEREO/WAVES near Earth's orbit   measured nanodust   with a variable flux much higher than that of larger particles \citep{mey09b,zas12,lec13}. Nanodust was also detected on Cassini/RPWS when the instrument was episodically turned on during  its cruise phase;  the flux was highly variable \citep{sch14} and decreased  as the inverse squared heliocentric distance \citep{sch15}.  Cassini observations during corotating fast and slow solar wind streams (CIR) confirmed that nanodust particles  move at speeds near  $v_D$ \citep{mey16}. Cassini/RPWS measured also  jovian nanodust streams,  in agreement with dedicated dust detectors \citep{mey09a}. 

\begin{figure*}
	\centering
	\includegraphics[width=12cm]{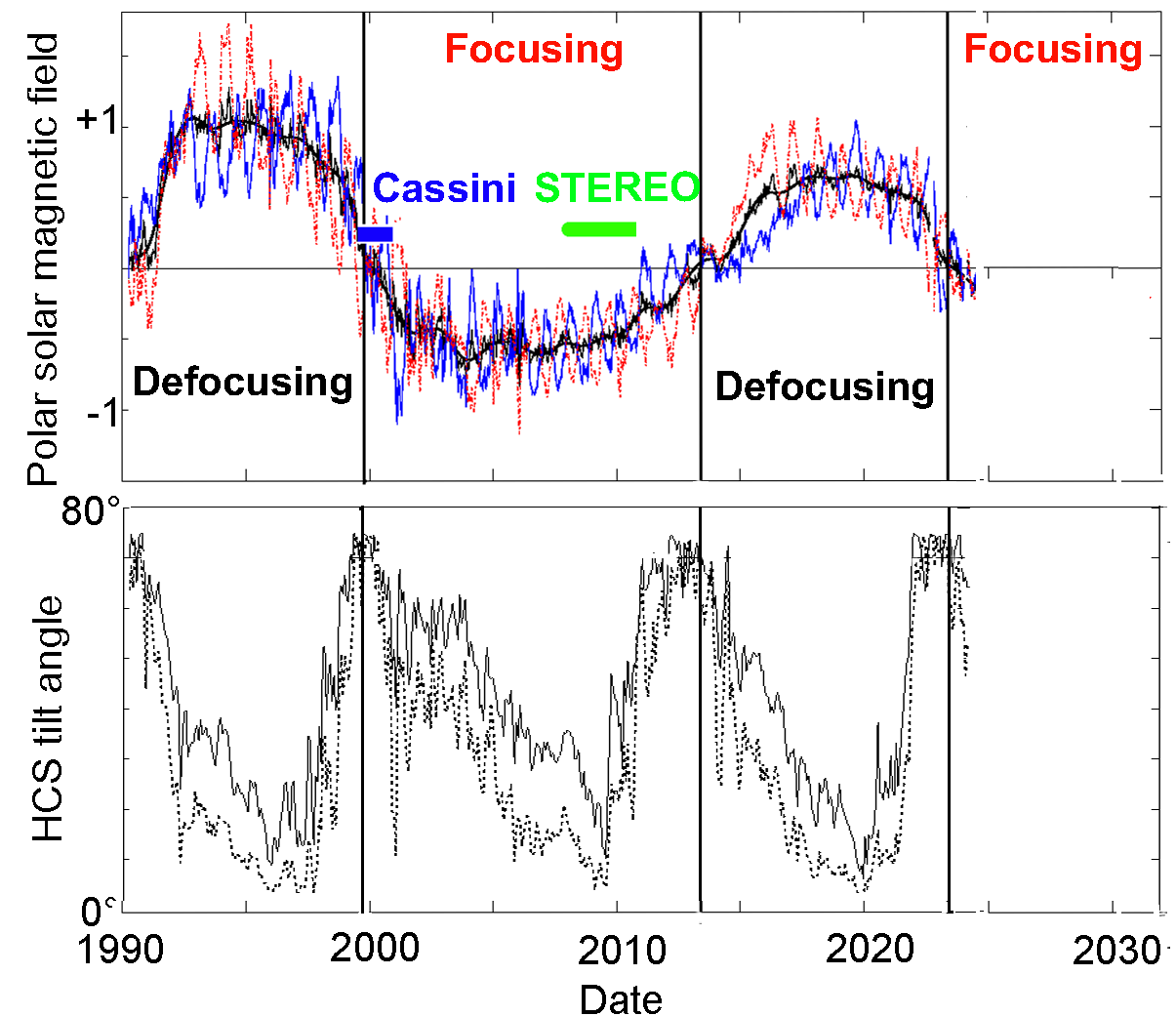}
	\caption{
	Top. Polar solar magnetic field strength (in Gauss), from the Wilcox Solar Observatory (+ North: solid blue, - South: dashed red, average: medium  solid, smoothed average: heavy solid).  Negative (positive) north magnetic field produce a focusing (defocusing) electric field for positive charges. Both STEREO and Cassini nanodust observations took place during a focusing configuration. Bottom. Maximal tilt angle of the HCS (solid: classic PFSS model, dashed: radial boundary condition model). \label{Fig1}}
\end{figure*}

Nanodust was much easier to detect on Cassini than on STEREO because the power spectra are  proportional to the impact rate whereas Cassini spacecraft projected surface  area exceeded that of STEREO by one order of magnitude. Furthermore, the    waveform measurement on Cassini had a higher sensitivity than on STEREO.

The high variability of the interplanetary nanodust  flux observed by  STEREO  and Cassini  was attributed to time-dependent production, transient magnetic field changes, and to the complex  dynamics in presence of the HCS  \citep{cze12,juh13,pop22}. A variable production rate was confirmed by  PSP dust measurements \citep{sza21};  furthermore,  even if nanodust was uniformly distributed near the Sun,  dynamics would produce a highly nonuniform flux near  Earth's orbit.

An important parameter governing nanodust detectability by STEREO and Cassini   is the polarity of the solar magnetic dipole  \citep{mey07}, which determines whether the   interplanetary electric field is focusing or defocusing  \citep{juh13,pop22}. A defocusing electric field points away from the HCS, driving  positively charged nanodust away from it; larger particles, having a faster vertical drift, quickly drift away; hence, only the smallest particles  can go close to it and be observed near the ecliptic. In contrast, in focusing conditions, the larger nanograins tend to move near the HCS, enabling a broader size range to be detected when the HCS is weakly  tilted and thus close to the ecliptic.

Figure 1 shows the periods   of nanodust observation on STEREO and Cassini. Both occured during focusing conditions.

However,  the periods of nanodust observation on Cassini  (from 01:1999 to 03:2000) were  determined by the instrument itself, which was  turned on only episodically; furthermore,  beyond the asteroid belt,  the integration time  was considerably reduced  in order to measure  radioemissions at a high rate, making the instrument  unable  to detect dust \citep{sch15}.   Cassini nanodust detection  took place near solar activity maximum, so the HCS was  generally far from ecliptic (see Fig. 1), which explains  that the detected particles  were  smaller than 10 nm (see Fig 2 in \citet{mey16}), since these small particles have a smaller vertical drift  \citep{juh13,pop22}.

Interplanetary nanodust observations on STEREO started at the beginning of the mission and ended before the end of the focusing period, when    the  tilt angle of the HCS became large (see Fig.1). This indicates  that, as suggested by \citet{pop22}, STEREO  A could only measure   the largest nano grains ($>10$ nm), which are most sensitive to it and were far from ecliptic when the HCS tilt was large. STEREO measuring larger nanograins than Cassini is coherent with the already mentioned higher sensitivity of Cassini for nanodust detection.

\section{Concluding remarks}
Cassini observed  nanodust   when the HCS was highly tilted, but it  observed small nanodust particles, which are weakly affected by the vertical drift. In contrast, STEREO  observed  nanodust when the  focusing interplanetary electric field pushed large nano particles towards the HCS, but the detection stopped when the HCS tilt angle increased too much.    This indicates that nanodust may reappear in STEREO wave data, as suggested by  \citet{pop22}, when the  HCS tilt is not too high.

Parker Solar Probe and Solar Orbiter orbit too close to the Sun for the  speed of nanodust to be large enough to enable its detection by wave instruments, except when transients, as  coronal mass ejections, would accelerate it sufficiently \citep{lec15,obr18}. Such events would open up the possibility of nanodust detection at small heliocentric distances.

\vspace{1cm}
%\begin{acknowledgments}
	The magnetic field data were obtained at \url{http://wso.stanford.edu/}.
%\end{acknowledgments}	 

\end{document}